\documentclass[reprint,aps,prl,twocolumn,groupedaddress,amsmath,amssymb]{revtex4-1}
\usepackage{graphicx}  % needed for figures
\usepackage{dcolumn}   % needed for some tables
\usepackage{bm}        % for math
\usepackage{verbatim}   % for math
\usepackage[mathscr]{euscript}
\usepackage[font=small,skip=1pt]{caption}
\usepackage{epstopdf}
\usepackage{caption}
\usepackage{ragged2e}
\usepackage{xcolor}
\newcommand{\bea}{\begin{eqnarray}}
\newcommand{\eea}{\end{eqnarray}}

\newcommand{\ket}[1]{|#1\rangle}
\newcommand{\bra}[1]{\langle#1|}
\newcommand{\dn}{{\downarrow}}
\newcommand{\up}{{\uparrow}}

\begin{document}

\title{Discrete fluctuations in memory erasure without energy cost}
\author{{Toshio} Croucher}
\affiliation{
   Centre for Quantum Dynamics,\\
   Griffith University,\\
   Brisbane, QLD 4111 Australia
   }
\author{Salil Bedkihal}
\affiliation{
   Centre for Quantum Dynamics,\\
   Griffith University,\\
   Brisbane, QLD 4111 Australia
   }

   \author{Joan A. Vaccaro}
\affiliation{
   Centre for Quantum Dynamics,\\
   Griffith University,\\
   Brisbane, QLD 4111 Australia
   }
\date{\today}
\begin{abstract}
According to Landauer's principle, erasing one bit of information incurs a minimum energy cost. Recently, Vaccaro and Barnett (VB) explored information erasure within the context of generalized Gibbs ensembles and demonstrated that for energy-degenerate spin reservoirs, the cost of erasure can be solely in terms of a minimum amount of spin angular momentum and no energy.
As opposed to the Landauer case, the cost of erasure in this case is associated with the discrete variable.
 Here we study the {\it discrete} fluctuations in this cost and the probability of violation of the VB bound. We also obtain a Jarzynski-like equality for the VB erasure protocol.  We find that the fluctuations below the VB bound are exponentially suppressed at a far greater rate and more tightly than for an equivalent Jarzynski expression for VB erasure. We expose a trade-off between the size of the fluctuations and the cost of erasure. We find that the discrete nature of the fluctuations is pronounced in the regime where reservoir spins are maximally polarized.
We also state the first laws of thermodynamics corresponding to the conservation of spin angular momentum for this particular erasure protocol.
Our work will be important for novel heat engines based on information erasure schemes that do not incur an energy cost.
\end{abstract}

\maketitle
Understanding the thermodynamical costs of information erasure \cite{Szilard1929,Landauer1961} is of fundamental importance for the design of nanoscale heat engines \cite{Dechant2015} and reversible computing \cite{Levitin2007, Vos2011}.
Landauer's erasure principle expresses a fundamental bound for any erasure process that uses a thermal reservoir to store the erased information. \cite{Plenio01thephysics}. It states that the work cost $W$ to erase one bit of information \cite{Landauer1961} is given by
\vspace{-2mm}
\bea
W \geq \beta^{-1}\ln 2 \label{eqn:1}
\eea
where $\beta=1/k_{B}T$, $T$ is the temperature of the reservoir and $k_B$ is Boltzmann's constant.
Recently, Dillenschneider and Lutz \cite{Dillenschneider2009} generalized Landauer's principle to accomodate fluctuations and obtained the probability of violating Eq. (\ref{eqn:1}) as
\bea
P(W\le\beta^{-1}\ln 2-\epsilon)\leq e^{-\beta\epsilon}. \label{eqn:2}
\eea
In other words, small fluctuations $\epsilon$ below the Landauer bound are possible and the probability of violation is exponentially suppressed in accordance with the Jarzynski equality \cite{Jarzynski1997, Sagawa2010}.

The association between information erasure and energy embodied in Eq. (1) has been widely accepted as a natural one.
This is perhaps due to the deep connection between energy and entropy in traditional thermodynamics. However, in two classic papers, Jaynes \cite{Jaynes1957, Jaynes1957b} formulated a generalized theory of statistical mechanics using a {\it principle of maximum entropy} where not only energy but all other {\it measurable conserved} quantities can be treated on an equal footing.
In this framework the notion of heat can be generalized to incorporate an exchange of arbitrary conserved quantities such as quantized spin angular momentum. In other words if there are $k$ conserved quantities associated with the generalized reservoir, then the corresponding heat is called the ``$k^{th}$" heat and the corresponding measurement probe is called the ``$k^{th}$" meter \cite{Jaynes1957, Jaynes1957b}. If the $k^{th}$ conserved quantity corresponds to energy then the corresponding meter is the thermometer.

Recently, Vaccaro and Barnett  (VB) \cite{Vaccaro2011,Barnett2013} applied Jaynes' framework to the problem of erasing information when multiple quantities are conserved.  In particular, they formulated a protocol based on energy-degenerate spin-$\frac{1}{2}$ reservoirs under the conservation of spin angular momentum that gives the cost of erasure solely in terms of spin angular momentum as
\bea
     \mathcal{L}_{\rm s} \geq \gamma^{-1}\ln{2}. \label{eqn:VB}
\eea
Here $\mathcal{L}_{\rm s}$ is the spin equivalent of work which, henceforth, we refer to it as a {\it spinlabor} \footnote{ We introduce the composite words spinlabor and spintherm to describe the spin equivalent of work and heat based on the Latin word labor for work and the Greek word thermosh for warm $\theta \epsilon \rho \mu$ \'{o} $\zeta$} and $\gamma$ is the Lagrange multiplier associated with the conservation of spin angular momentum which can be expressed as \cite{Vaccaro2011,Barnett2013}
\bea
      \gamma=\frac{1}{\hbar}\ln\left[\frac{N\hbar-2\langle \hat{J}^{(R)}_{z}\rangle}{N\hbar+2\langle \hat{J}^{(R)}_{z}\rangle}\right]=\frac{1}{\hbar}\ln\left[\frac{1-\alpha}{\alpha}\right]\label{eqn:gamma}
\eea
where $\langle\hat{J}^{(R)}_z\rangle=\left(\alpha-\frac{1}{2}\right)N\hbar$ is the $z$ component of the total spin angular momentum, $N$ is the number of spins in the reservoir and $0\le \alpha \le 1$ represents a convenient spin polarisation parameter.
It is interesting to note that the Boltzmann constant does not appear in Eq. \eqref{eqn:gamma} since there is no {\it energy cost} involved due to the {\it degeneracy} of the internal (intrinsic spin) degree of freedom. In other words, the spatial and spin degrees of freedom are decoupled from each other, a situation that is realized in optical trapping of cold atomic gases \cite{Leanhardt2003}. Also note that $\gamma$ in Eq. \eqref{eqn:gamma} has been called the inverse ``spin temperature'' in optical pumping where spin exchange collisions lead to the redistribution of spin angular momentum \cite{Anderson1961,Appelet1998}. %This notion of spin temperature is different than one used in solid state physics where spin degree of freedom is assumed to be equiliberated with a lattice temperature \cite{Abragam1958}.
%If spin degeneracy in the reservoir is broken by Zeeman field (symmetry breaking) then coupling of spatial and spin degrees of freedom will result in an erasure with multiple costs involving both energy and loss of spin polarization \cite{Vaccaro2011,Barnett2013}.
If the energy degeneracy is broken by a Zeeman field then the coupling of spatial and spin degrees of freedom will result in an erasure with multiple costs in terms of both energy and spin polarization \cite{Vaccaro2011,Barnett2013}.
Very recently, Jaynes' framework has been further extended to incorporate non-commuting degrees of freedom, and this led to a novel classification of Gibbs  states, namely Abelian and non-Abelian thermal states, and the possibility of extracting arbitrary conserved quantities from individual quantum systems \cite{PST11, Halpern2015,Lostaglio2015}.

In this work we consider a collection of energy-degenerate spins as an explicit physical example of a generalised reservoir in Jaynes' formalism, and study the discrete fluctuations in an energy-free erasure scheme \cite{Vaccaro2011,Barnett2013}, a scenario that remains unexplored.
%It differs from previous studies of fluctuations which mainly focused on thermodynamics of systems involving continuous degrees of freedom.
Our main results are a Jarzynski-like equality for quantized spin angular momentum exchange, and the probability of violation of the VB bound given by Eq. \eqref{eqn:VB}.
We find that fluctuations below the VB bound are suppressed at a rate far greater than expected from the equivalent Jarzynski equality.

We further find that the effects of discreteness are more pronounced at higher degrees of reservoir spin polarization $\alpha\to 0$  or equivalently, from  Eq. \eqref{eqn:gamma}, at low spin temperatures $\gamma^{-1}\to 0$.
This is the first time that the fluctuations in the memory erasure has been studied for discrete system. Our work also opens up the possibilities of novel fluctuation theorems for other conserved quantities, especially in the context of generalized Gibbs ensembles, e.g. integrable spin chains  \cite{Balazs2013}.
%Our work will be important for novel heat engines based on generalized reservoirs with arbitrary conserved quantities.

\vspace{-5mm}
\section{The Model}

We briefly outline the VB erasure scheme here---full details of can be found in Ref. \cite{Vaccaro2011,Barnett2013}.  Studies of information erasure typically involve two-state memory systems in contact with one or more reservoirs.  In the VB scheme the memory logic states are associated with the $z$ component of spin polarization with the eigenstate $|{\downarrow}\rangle$ corresponding to eigenvalue $-\hbar/2$ representing logical $0$ and $|{\uparrow}\rangle$ corresponding to $\hbar/2$ representing logical $1$. These states are assumed to be energy degenerate so that the erasure incurs no energy cost.
The reservoir that acts as an entropy sink consists of $N$ similar energy-degenerate spins.
The only other information we have about the reservoir is the expectation value of the $z$ component
of spin polarization $\langle\hat{J}^{(R)}_{z} \rangle$ and the average energy associated with the motional degrees
of freedom $\langle\hat{H}^{(R)}_{\rm ext}\rangle$. According to the {\it maximum entropy principle} \cite{Jaynes1957},
the best description of the reservoir is then given by the density operator \cite{Barnett2013}
\bea
\hat{\rho}=\frac{\exp(-\beta\hat{H}^{(R)}_{\rm ext}-\gamma \hat{J}^{(R)}_z)}{Z} \label{eqn:density op}
\eea
where $\beta$ and $\gamma$ are corresponding Lagrange multipliers which are independent of each other, and $Z$ is the partition function.
%In our case, because of the energy {\it degeneracy} the information  erasure occurs by spin angular momentum  exchange only. %In this particular case of erasure, the internal energy of each spin is constant and therefore the entropy associated with spatial degrees of freedom remains constant, and the entropy associated with spin degree of freedom plays a role.
Such density operators may be realized in dilute cold atomic gases confined in optical traps \cite{Leanhardt2003} and for certain classes of integrable spin chain models \cite{Balazs2013}.

We assume the reservoir is sufficiently large (i.e. $N\gg 1$) so that erasing one bit of information changes $\gamma$ by a negligible amount and so the spin angular momentum of the reservoir will be described by an approximately-fixed probability distribution; specifically
\bea
P_{\uparrow}(n,\nu)=\frac{\exp(-\gamma n\hbar)}{Z_R}, \label{eqn:probr}
\vspace{-2mm}
\eea
is the probability that the reservoir has a $z$ component of spin polarization of $(n-\frac{1}{2}N)\hbar$ and the arrangement $\nu$  where $\nu=1, 2,\ldots {}^{N}C_{n}$ indexes a unique arrangement of the $N$ individual spin states and $Z_{R}$ is the associated partition function.
We also assume that the memory spin is in an arbitrary state initially.
It should be noted that due to the energy degeneracy of the spins in the reservoir and memory, we only need to consider spin angular momentum exchange. We allow energy flow between spatial degrees of freedom, but it does not contribute to the erasure cost.

Following the protocol in Refs. \cite{Vaccaro2011,Barnett2013}, we make use of an energy degenerate ancillary spin-$\frac{1}{2}$ particle that is initially in a state $|{\downarrow}\rangle\langle{\downarrow}|$ corresponding to logical zero.
A controlled-not (CNOT) operation is then applied to the memory-ancilla system with the memory spin acting as the control and the ancilla spin as the target. The spinlabor cost for this initial step is $\frac{\hbar}{2}$ and it leaves both memory and ancilla spins in an equal mixture of both spin up and spin down.
After the CNOT operation  we allow the memory-ancilla system to reach spin equilibrium with the reservoir by the particular exchange of spin angular momentum according to the protocol devised by VB \cite{Vaccaro2011,Barnett2013}. This equilibration step is assumed to conserve the total spin angular momentum.
Cycles consisting of adding an additional ancilla to the memory-ancilla system, the CNOT operation, and spin equilibration with the reservoir are repeated until the desired degree of erasure is achieved.
%We define $P_{k}(i)$ as the probability of the CNOT operation cost for a particular state $i$. Here the state $i$ is referred to as the number of $\hbar$ spin angular momentum that has been used up for the $k^{th}$ iteration. For the first iteration we have
%$k=1$, we expect 2 outcomes, no cost in spin angular momentum or the $\hbar$ cost. The probability of CNOT cost is given by,
%\bea
%P_{\mbox{cost}}=P^{k}=\frac{\exp(-\gamma k\hbar)}{1+\exp(-\gamma k\hbar)} \label{eqn:1}
%\eea
%and for no cost,
%\bea
%P_{\mbox{no cost}}=1-P^{k}=\frac{1}{1+\exp(-\gamma k\hbar)}. \label{eqn:2}
%\eea
%So for the first iteration $P_{1}(0)=P_{\mbox{no cost}}$ and $P_{1}(1)=P_{\mbox{cost}}.$
%Each final state after an iteration has two possible outcomes, the first is that it remain in the current state given by the probability of $\eqref{eqn:1}$, and the second is that it moves to a state that costs $\hbar$, and is given by $\eqref{eqn:2}$. This can be expressed in terms of a recursive sequence as,
%\bea
%P_{k+1}(i)=P_{\mbox{no cost}}P_{k}(i)+P_{\mbox{cost}}P_{k}(i-1).
%\eea
%It should be noted that $P_{\mbox{no cost}}$ and $P_{\mbox{cost}}$ are probabilities for the $k+1^{th}$ iteration.
In each cycle, the CNOT operation incurs a cost of $\hbar$ in spin angular momentum in transforming the newly added ancilla from $\ket{\dn}\bra{\dn}$ to $\ket{\up}\bra{\up}$ if the memory spin is in the state $\ket{\up}\bra{\up}$. On average, the cost of the CNOT operation for the $m^{th}$ cycle is given by  $\hbar Q_\uparrow(m)$ where
\bea
      Q_\uparrow(m) =\frac{e^{-(m+1)\gamma\hbar}}{1+e^{-(m+1)\gamma\hbar}}  \label{eqn:Q}
\eea
is the probability of the memory spin being in $\ket{\up}\bra{\up}$.
The total cost $\mathcal{L}_{\rm s}=\frac{\hbar}{2}+\sum_{m=1}^{\infty} \hbar Q_\uparrow(m)$ is bounded below by Eq. \eqref{eqn:VB}.
Our objective is to study the fluctuations in this cost. % $\mathcal{L}_{\rm s}$.
% of the VB erasure.
The combined reservoir (R) and memory-ancilla (M) system is isolated except for times when the unitary CNOT operations take place. By conservation of spin angular momentum, any spinlabor performed by the CNOT operation will result in the change $\mathcal{L}_{\rm s}=\Delta J_{z}^{(T)}$ to the total spin angular momentum $J_{z}^{(T)}=J_{z}^{(R)}+J_{z}^{(M)}$.  Using Liouville's theorem, we first obtain \footnote{See supplementary material} the equivalent of Jarzynski's equality \cite{Jarzynski1997, Jarzynski1997a}
\bea  \label{eqn:A}
\langle e^{(-\gamma\mathcal{L}_{\rm s}+\ln2 )}\rangle &=& \sum_{\mathbf{z}}f(\mathbf{z})e^{[-\gamma\hbar(\Delta m^{(R)}_{j}+\Delta m^{(M)}_{j})+\ln 2]} \\
 &=& \frac{1+e^{-\gamma \hbar}}{1+e^{-2\gamma\hbar}}=A,\nonumber
\eea
where $f(\mathbf{z})$ is the probability distribution over phase space vectors $\mathbf{z}$ which indexes internal spin angular momentum eigenvalues $m_j^{(\cdot)}$ and external spatial coordinates. We define the probability that the cost in spinlabor be $\mathcal{L}_{s}$ as $Pr(\mathcal{L}_{s})\equiv f(\Delta J_{z}^{T}=\mathcal{L}_{s})$.
%\bea
%\langle \exp(-\gamma \Delta J_{z}-\Delta I) \rangle =1,
%\eea
%Using the probabilities defined in Eq. \eqref{eqn:probr} and Eq. \eqref{eqn:probm} we find
%\bea
%1 & =& \sum_{x,x'} P(x)P(x'|x) \times \nonumber \\
%&& \exp(-\gamma \Delta J_{z}-\Delta I) \nonumber \\
%&=& \sum_{x,x'} P(x)P(x'|x) \frac{P(n,\nu)}{P(n',\nu')} \frac{Q(m)}{Q(m')}\nonumber \\
%&=& \sum_{x,x'} P(x)P(x'|x) \frac{P(x)}{P(x')} \nonumber \\
%&=& \sum_{x,x'} P(x)P(x|x') =1
%\eea
Following Jarzynski's analysis \cite{Jarzynski2011} we find the probability that the spinlabor cost $\mathcal{L}_{s}$ violates VB's bound by $\epsilon$ satisfies
\bea
{Pr^{(v)}(\epsilon)}\equiv{Pr}(\mathcal{L}_{\rm s}\leq \gamma^{-1}\ln{2}-\epsilon ) \leq A e^{-\gamma\epsilon } \label{eqn:bound1},
\eea
where $A$ is given by Eq.~(\ref{eqn:A}). The above result is analogous to the Jarzynski's work on the probability of observing a violation of the Clausius-Duhem inequality  \cite{Jarzynski1999,Jarzynski2011}.
Judiciously restricting the sum on the right side of Eq.~(\ref{eqn:A}) yields a tighter bound on $Pr^{(v)}(\epsilon)$ \cite{Note2}
\bea
{Pr^{(v)}(\epsilon)}
%={Pr}(\mathcal{L}_{\rm s}\leq \gamma^{-1}\ln{2}-\epsilon )
\leq B e^{-\gamma\epsilon }    \label{eqn:bound2},
\eea
where $B=\sum_{\Delta J_{z}^{(T)}\leq \gamma^{-1}\ln 2} f(\mathbf{z})
   e^{[-\gamma\hbar(\Delta m^{(R)}_{j}+\Delta m^{(M)}_{j})+\ln 2]}$ and $B\le A$.
Using semi-analytic methods \cite{Note2} we find an even tighter bound for $\gamma\to 0$ (i.e. $\alpha\to 0.5$)
\bea
    {Pr^{(v)}(\epsilon)}
    %= {Pr}(\mathcal{L}_{\rm s}\leq \gamma^{-1}\ln{2}-\epsilon )
    \leq  C e^{-\sqrt{\frac{\gamma}{\hbar}}\epsilon}
     \label{eqn:bound3},
\eea
where $C=Pr(\mathcal{L}_{\rm s}\leq \gamma^{-1}\ln{2})$.
Eqs.~(8)-(11) are the central novel results of our work.
%The probability of violation can be extended for multiple bits when an erasure of each is independent of the other, and is given by
%\bea
%\Pi_{\;q}\left[Pr_{q}\left( \frac{\Delta J_{z}}{q}< -\gamma^{-1}\frac{\Delta I}{q}-\epsilon \right)\right] \leq e^{-\sqrt{q}\;\gamma\epsilon}, \label{eqn:genprobN}
%\eea
%where $q$ is the number of bits erased.
\begin{figure}[ht]
\centering
\captionsetup{justification=RaggedRight}
  \vspace{-5mm}
 \includegraphics[width=0.38\textwidth]{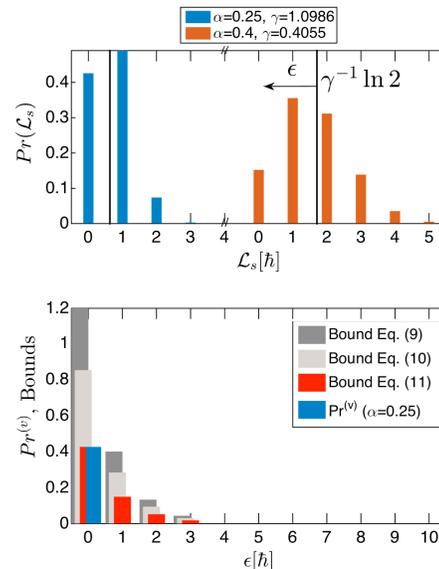}
  \vspace{-5mm}
  \caption{(a) Spinlabor statistics in the regime $\alpha\leq 0.4$. The vertical black line represents the bound $\gamma^{-1}\ln{2}$. (b) Comparison of bounds in Eq. \eqref{eqn:bound1}-\eqref{eqn:bound3} and the probability of violation.}
  \label{fig:low_alpha}
  \end{figure}

\begin{figure}[ht]
\centering
\captionsetup{justification=RaggedRight}
  \vspace{-5mm}
      \includegraphics[width=0.38\textwidth]{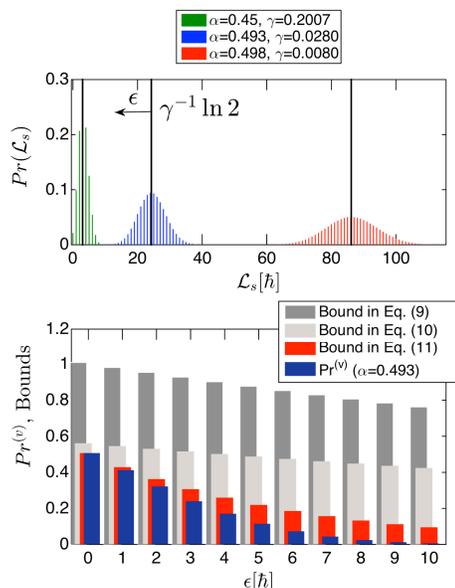}
  \vspace{-5mm}
  \caption{(a) Spinlabor statistics in the limit $\alpha>0.4$. Here the vertical black line represents the bound $\gamma^{-1}\ln{2}$. (b) Eq. \eqref{eqn:bound1}, Eq. \eqref{eqn:bound2} and Eq. \eqref{eqn:bound3} show an exponential suppression. Blue histogram shows probabilities below the Vaccaro-Barnett bound.}                                                                                                                                                                         
  \label{fig:high_alpha}
  \vspace{-5mm}
\end{figure}

We illustrate the spinlabor statistics and the probability of violation for two distinct regimes of the reservoir spin polarization: (1) $\alpha\leq 0.4$ and (2) $\alpha>0.4$ in Fig.~(1) and Fig.~(2) respectively.
Fig.~1(a) compares the probability of the spinlabor cost $Pr(\mathcal{L}_{\rm s})$ for two values of $\alpha\leq 0.4$ with the corresponding VB bounds $\gamma^{-1}\ln{2}$ (vertical black lines) and shows that there is a significant probability of violating the bound (bars left of the bounds).
This is analogous to the fluctuation in the work cost below Landauer's bound, as shown in Ref. \cite{Dillenschneider2009}. %The overlap of the probability distributions in Fig.1(a) can be attributed to almost identical values of $\gamma$ for chosen spin polarizations.
The probability of violation $Pr^{(v)}$ is represented in Fig.~1(b) which shows that the fluctuations are exponentially suppressed even faster than our tighter bounds in Eq. \eqref{eqn:bound2}. and Eq. \eqref{eqn:bound3}.

Fig.~2 shows the spinlabor statistics and the probability of violation for three different values of $\alpha$ in the regime $\alpha>0.4$.
As $\alpha$ increases, Fig.~2(a) shows  the probability distribution $Pr(\mathcal{L}_{\rm s})$ tends to become symmetric.  Correspondingly, $\gamma \to 0$, the reservoir approaches an equal mixture of spin up and spin down and its entropy approaches its maximum value.  This leads to increasing fluctuations in the erasure which is reflected in $Pr(\mathcal{L}_{\rm s})$ becoming broader in absolute terms.  Nevertheless, the fluctuations relative to the erasure cost become narrower and the cost diverges, as can be seen from Eq. \eqref{eqn:VB}.

Figures 1(a) and 2(a) also show a tradeoff between the average cost and the relative size of the fluctuations compared to the cost, as follows.  For $\alpha \leq 0.4$ the average cost is lower and the fluctuations are relatively pronounced, whereas for $\alpha>0.4$ the average cost is higher and the fluctuations relative to it are lower.

It is interesting to recast the task of erasing information in terms of spatial orientating.  A reservoir with polarised spins (i.e. $\alpha<0.5$) breaks rotational symmetry and becomes a resource for orientating the memory spin.  As $\alpha \to 0$, the polarisation increases and the reservoir becomes more asymmetric; correspondingly, the orientating task becomes easier and so the associated cost, $\mathcal{L}_{\rm s}$, lowers.  For the opposite trend, as $\alpha\to0.5$, the polarisation reduces and the reservoir becomes more rotationally symmetric.  This makes the orientating task increasingly difficult and so it incurs a rising cost.
%This is also reflected in the probability of violation as shown in Fig.~2(b), where the number of discrete jumps increase as $\alpha \to 0.5$, but the step size is reduced and the curve becomes quasi-continuous.
%Note that the averages of $Pr^{(cost)}(\Delta J_{z})$ are larger than the bound $\gamma^{-1}\ln{2}$ because the numerical calculations are exact whereas the bound is an approximation \cite{Vaccaro2011,Barnett2013}.

\vspace{-4mm}
\section{Discussion}
\vspace{-4mm}

Landauer's erasure principle has been shown to be equivalent to the second law of thermodynamics \cite{Plenio01thephysics}. Similarly, Vaccaro and Barnett's erasure scheme gives an illustrative example of a generalized form of the second law of thermodynamics for systems that exchange arbitrary conserved quantities.
Landauer's erasure principle has been shown to require modification for small systems where fluctuations are important \cite{Dillenschneider2009}.
In this work we explored the corresponding fluctuations in  Vaccaro and Barnett's erasure scheme. We showed that the
discrete nature of spin angular momentum exchange is reflected in the spinlabor statistics and the probability of violation. Although we analyzed an energy degenerate spin-$\frac{1}{2}$ systems in this work, our results can be extended to energy-degenerate arbitrary-spin angular momentum reservoirs.

In addition to having an impact for a generalized second law, this work has also implications for the first law as follows.
%The average total spin polarization $\ip{\hat{J}_{z}^{(T)}}=\ip{\hat{J}_{z}^{(R)}}+\ip{\hat{J}_{z}^{(M)}}$ of the reservoir (R) and memory-ancilla (M) system after the $m$th cycle of the erasure scheme.
In general, the average spin angular momentum can be written as
\bea
 {J}_{z}=\sum_{j,m_j} \hbar p(j,m_j)g(m_j)
\eea
where $p(j,m_j)$ is the probability associated with the spin state $(j,m_{j})$ and $g(m_j)=m_j$ initially. The total change in ${J}_{z}$ is given by \cite{Jaynes1957}
%First the CNOT operation increases the value of $m$ in the third term on the right by 1;
%this change represent an increase in the accumulated spinlabor cost $\mathcal{L}_{\rm s}$.
%Next, bringing the memory-ancilla system into spin equilibrium with the spin reservoir increases the value of $S$ in the first term;
%this change represents the transfer of spintherm (the spin equivalent of heat) to the reservoir.
%If we consider the second term to be unchanged, the change in $ \ip{\hat{J}_{z}^{(T)}}$ is
 \bea
\Delta{J_{z}}=\mathcal{L}_{s}+\mathcal{Q}_{s},
\label{eqn:1st law spin}
\eea
where  $\mathcal{L}_{s}=\sum_{j,m_j}\hbar p(j,m_j)\Delta g(m_j) $ is the part of the change that is deterministic in origin, e.g. due to unitary evolution by an external device, and $\mathcal{Q}_{s}=\sum_{j,m_j}\hbar g(m_j)\Delta p(j,m_j)$ is the part that is non-deterministic, e.g. due to the equilibration with another spin system. Eq.~(\ref{eqn:1st law spin}) is the spin equivalent of the first law
\bea
 \Delta U=W+Q.\label{eqn:1st law energy}
\eea
Indeed, we refer to $\mathcal{Q}_{s}$ as spintherm (i.e. as the spin equivalent of heat) in analogy with standard thermodynamics where a change in the density matrix is associated with heat exchange.
%where $\Delta J_{z}^{(T)}=\Delta \ip{\hat{J}_{z}^{(T)}}$ represents change in spin polarization, $\Delta S/\gamma$ represents spintherm, and $\Delta\mathcal{L}_{\rm s}=\hbar\Delta[{m}Q_\uparrow(m)]$ represents extra spinlabor.  Eq. \eqref{eqn:1st law spin} thus represents the first law for spin angular momentum.
For the particular case studied here, where one bit of information is erased from a spin-$\frac{1}{2}$ memory, the change in the spin angular momentum of the memory is $\Delta{J_{z}^{(M)}}=-\frac{1}{2}\hbar$, and so according to Eq.~(\ref{eqn:1st law spin})  
\bea    \label{eqn:spintherm removed}
  -\mathcal{Q}_{s}= \mathcal{L}_{s}+\frac{1}{2}\hbar.
\eea
%The above inequality shows that, unlike the Landauer case where heat and work costs are equivalent, in this case 
The total spintherm is greater than the spinlabor cost because the memory initially contains inherent spintherm of $\frac{1}{2}\hbar$; this contrasts with the usual case in Landauer's erasure where heat and work costs are equivalent.
%This follows from the discreteness of the spin angular momentum.
%In general the equivalence between ``$k^{th}$" heat and ``$k^{th}$" work need not hold for other other conserved quantities. In this particular case, the spin angular momentum plays the role of internal energy, but unlike the traditional thermodynamics where $\Delta U$ may be zero, the change in spin angular momentum is always non-zero which leads to this inequivalence of spinlabor and spintherm.
%Also the negative sign of spintherm implies loss and therefore entropy in the memory is dumped to the spin reservoir. 
The negative sign in the expression $-\mathcal{Q}_{s}$ implies that the spintherm is removed from the system.
In this sense the VB erasure protocol is an example of a \emph{spintherm pump}.
%The principles that led to Eq. \eqref{eqn:1st law spin} also allow a differential form of the first law as
% $d{J_{z}^{(T)}}=\delta \mathcal{Q}_{\rm spin}+\delta\mathcal{L}_{\rm s}$
%where $\delta\mathcal{Q}_{\rm spin}={\rm Tr}[\hat{J}^{(R)}_{z}d_{t}\rho]$ is the path-dependent spintherm and $\delta\mathcal{L}_{\rm s}={\rm Tr}[\rho(t)d_{t}\hat{J}^{(M)}_{z}]$ is the path-dependent spinlabor. \textbf{This could be deleted because it contains things that are not described (e.g. path-dependent).}

Moreover, the number of first laws of generalized thermodynamics are equal to number of conserved quantities if all the Lagrange multipliers are independent of each other. In our case we have two independent laws: Eq. \eqref{eqn:1st law energy} which corresponds to the conservation of energy in the energy exchange between the spatial degrees of freedom (that does not contribute to the cost of the erasure), and Eq. \eqref{eqn:1st law spin} which corresponds to the conservation of intrinsic spin angular momentum.

%If these degrees of freedom are mixed (such as the case of adiabatic demagnetization where spin degree of freedom is also assumed to have the same inverse temperature as that of the lattice) then one gets the usual continuous differential form of the first law.
Finally, we should mention that the experimental realisation and manipulation of spin reservoirs is not new. For example,  optical pumping is a well established method to create spin polarized gases \cite{Anderson1961, Appelet1998}.  In particular, in spin-exchange optical pumping of ${}^{3}{\rm He}$, the spin polarization is transferred between alkali atoms of certain polarization and the ${}^{3}{\rm He}$ nuclei \cite{Appelet1998}.
The relaxation processes here include spin exchange collisions between alkali atoms and the ${}^{3}{\rm He}$ nuclei. This is an example of entropy erasure by spin exchange that does not appear to have been appreciated, especially in the context of information erasure and thermodynamics.

To conclude, in contrast to recent work on abstract conserved quantities in resource theories using a generalised Gibbs state formalism \cite{PST11, Halpern2015, Lostaglio2015},
we have considered Vaccaro and Barnett's (VB) erasure scheme \cite{Vaccaro2011,Barnett2013} in which the conserved quantity is the physically-important spin angular momentum.  This observable has the distinction of having eigenvalues of integer multiples of $\hbar$. By studying the discrete fluctuations in VB memory erasure scheme, we exposed several interesting features; (a) a cross-over from the discrete to quasi-continuous spinlabor probability distribution in the limit when $\alpha\to0.5$, (b) the trade-off between the cost of erasure and the relative fluctuations, (c) a faster suppression (faster than Jarzynski's analysis) of the probability of violating VB bound (at a rate $\sqrt{\gamma/\hbar}$ found from the semi-analytic calculation).
 Such a faster suppression than Jarzynski's analysis is the novel result of our work and has not been reported before in the context of the fluctuations in memory erasure. Our work opens up an interesting possibility of flucutation relations within the context of generalized-Gibbs ensembles. %Future work will be devoted to explore the trade-off between fluctuations when multiple conserved quantities are involved in the erasing process.
%We analysed the associated discrete fluctuations in the cost of erasure. We derived a Jarzynski-like equality and the bound it gives for fluctuations that violate VB's minimum cost.
%However, this bound was found to be relatively weak and tighter bounds with greater suppression of the violation were obtained.
%The actual probability of violation is suppressed even faster than the bounds obtained.
%These results advance the study of novel heat engines and heat pumps that arise for generalised Gibbs ensembles.
%, such as ones running between thermal and energy-degenerate spin reservoirs \cite{Vaccaro2011,Barnett2013}.
%However, we found numerically that the probability of violation was suppressed at a rate that is even more rapid than exponential.
%We also exposed the tradeoff between the cost of erasure and relative size of the fluctuations where we found essentially that changing one inversely affects the other.

%We conclude our analysis by stating the first law of thermodynamics corresponding to the conservation of spin angular momentum in Eq. \eqref{eqn:1st law spin}.
%The fact that multiple conserved quantities, and their associated first laws, can act simultaneously has been shown here to open up possibilities fornovel heat engines such as ones running between thermal and energy degenerate spin reservoirs, and ``$k^{th}$" heat pump using generalized Gibbs ensembles.

\section{Acknowledgements}
This research was supported by the ARC Centre of Excellence Grant No. CE110001027, ARC Linkage Grant No. LP140100797 and the Lockheed Martin Corporation.  JAV thanks S.M. Barnett, J. Jeffers, J.D. Cresser and B. Schumacher for helpful discussions.

%\begin{figure}
%\begin{minipage}[t]{0.48\linewidth}
%\includegraphics[width=.9\linewidth]{circuits.png}
%\caption{This is a long sentence to test whether the caption is justified
 %or not. Unfortunately it is not.}
%\end{minipage}\hfill

\bibliography{paper_Edited}
\end{document}

\title{Discrete fluctuations in memory erasure without energy cost - Supplementary Material}
\author{{Toshio} Croucher}
\affiliation{
   Centre for Quantum Dynamics,\\
   Griffith University,\\
   Brisbane, QLD 4111 Australia
   }
\author{Salil Bedkihal}
\affiliation{
   Centre for Quantum Dynamics,\\
   Griffith University,\\
   Brisbane, QLD 4111 Australia
   }

   \author{Joan A. Vaccaro}
\affiliation{
   Centre for Quantum Dynamics,\\
   Griffith University,\\
   Brisbane, QLD 4111 Australia
   }
\date{\today}
\maketitle
\section{Recurrence relation}
In this section we derive a recurrence relation that is used to obtain spinlabor statistics in Fig.~(1) and Fig.~(2) in the main text.
We begin by briefly reviewing the Vaccaro-Barnett (VB) scheme.  The memory logic states in the scheme are associated with the $z$ component of spin polarization with the eigenstate $|{\downarrow}\rangle$ corresponding to eigenvalue $-\hbar/2$ representing logical $0$ and $|{\uparrow}\rangle$ corresponding to $\hbar/2$ representing logical $1$. These states are assumed to be energy degenerate so that the erasure incurs no energy cost.
The reservoir that acts as an entropy sink which consists of $N$ similar energy-degenerate spins in the state described by Eq.~(5) in the main text.
The erasure proceeds in two steps: (a) The memory spin is combined with an energy-degenerate ancilla spin that is initially in the state $|{\downarrow}\rangle$ to form the memory-ancilla system.
A controlled-not (CNOT) operation is applied to the memory-ancilla system with the memory spin acting as the control and the ancilla spin as the target.
(b) After the CNOT operation is applied we allow the memory-ancilla system to reach spin equilibrium with the reservoir by a particular exchange of spin angular momentum as discussed in the main text.
A cycle consisting of adding an extra ancilla to the memory-ancilla system, a CNOT operation with the memory spin as the control and the newly-added ancilla spin as the target, and spin equilibration with the reservoir is repeated until the desired degree of erasure is achieved.

%We assume that the memory spin is in a maximally mixed state initially.
Consider the situation at the end of the $m^{\rm th}$ cycle when the memory-ancilla system contains $m$ ancilla spins.
The memory-ancilla system has just undergone equilibration with the spin reservoir and this ensures the probability that the memory spin (and, correspondingly, all the ancilla spins) is in the state $|{\uparrow}\rangle$ is given by
\bea
Q_{\uparrow}(m)= \frac{e^{-(m+1)\gamma \hbar}}{1+e^{-(m+1)\gamma \hbar}}\label{eqn:1}
\eea
and in the state $|{\downarrow}\rangle$ by
\bea
Q_{\downarrow}(m)=1-Q_{\uparrow}(m)=\frac{1}{1+e^{-(m+1)\gamma \hbar}} \label{eqn:2}.
\eea
Let the probability that the CNOT operations have incurred a total cost of  $n\hbar$ over all $m$ cycles to this point be defined as $P_m(n)$, where  $0 \leq n \leq m$ because the cost is one $\hbar$, at most, each cycle.

In the subsequent cycle, the probability that the CNOT operation results in the newly-added ancilla changing from $|{\downarrow}\rangle$ to $|{\uparrow}\rangle$ and incurring a cost of one $\hbar$ is just $Q_{\uparrow}(m)$, and conversely, the probability that it results in no change and a zero cost is $Q_{\downarrow}(m+1)$.
There are two distinct ways in which the total cost at the end of the $m^{\rm th}$ cycle is $n\hbar$: at the beginning of the cycle either the total cost was $n\hbar$ and the memory spin was in the $|{\downarrow}\rangle$ state, or the total cost was $(n-1)\hbar$ and the memory spin was in the $|{\uparrow}\rangle$ state.
This leads to the following recurrence relation for the probability of the spinlabor cost:
\bea
P_{m+1}(n)&=&Q_{\downarrow} (m+1)P_{m}(n)\nonumber\\&&\quad +Q_{\uparrow} (m+1)P_{m}(n-1). \label{eqn:recrel}
\eea
We find the analytical solution to be
\begin{align*}
   P_{m}(q) = \frac{1}{\prod_{k=2}^{m}(1+r^{k})}
    \left\{
    \begin{array}{ll}
       \Lambda, & q\geq 1\\
       1, & q=0
    \end{array}
    \right.
\end{align*}
where
\begin{align*}
       \Lambda&=\left(\prod_{j=1}^{q}\frac{r^{j+1}-r^{m+1}}{1-r^{j}}\right) \left({1-p+p}\frac{1-r^{q}}{r^{q+1}-r^{m+1}}\right) ,\\
\end{align*}
$p$ is the initial probability that the memory spin is in the $\ket{\up}\bra{\up}$ state and
\begin{align}  \label{eqn:define r}
      r=e^{-\gamma \hbar} .
\end{align}
As the above solution can be verified by substitution into the recurrence relation \eq{eqn:recrel}, we omit the details of its derivation here.
The limit $m\to\infty$ of $P_m(q)$ corresponds to a \emph{full} erasure process.
Of particular interest is the result corresponding to the memory spin initially in an equal mixture of $\ket{\up}\bra{\up}$ and $\ket{\dn}\bra{\dn}$, i.e. $p=\frac{1}{2}$, in which case
\begin{align}
   P_{\infty}(q) &= \frac{1}{\prod_{k=2}^{\infty}(1+r^{k})}\non\\
    &\quad\times\left\{
    \begin{array}{ll}
       \left(\prod_{j=1}^{q}\frac{r^{j}}{1-r^{j}}\right)
       \frac{1-r^{q}(1-r)}{r} , & q\geq 1\\
       1, & q=0
    \end{array}
    \right.  \label{eqn:analy}
\end{align}
This is the probability that the spinlabor cost $\mathcal{L}_s$ is $q\hbar$ in a full erasure process.

%-----------------------------

\section{Jarzynski-like equality}
In this section we derive the Jarzynski-like equality given in Eq.~(8) of the main text.
We treat the spin reservoir and memory-ancilla as an isolated closed system undergoing deterministic evolution. The reservoir and memory system needs to be brought in to spin equilibrium by exchanging spin angular momentum.
This entails the spins in the reservoir plus memory-ancilla system to exchange internal spin angular momentum through elastic ``collisions'' of some kind involving the external degree of freedom. We assume that the reservoir-memory-ancilla system is isolated, which allows us to use Liouville's theorem in the following way.  First we assume that the reservoir-memory-ancilla system is described by the generalised Gibbs ensemble
\bea
f(\mathbf{z},t)=\frac{e^{(-\beta H_{\rm ext}^{(T)}-\gamma\hbar m_{j}^{(R)}-\lambda\hbar m_{j}^{(M)})}}{Z_{I} Z_{E}},
\eea
where $\mathbf{z}\equiv(j^{(R)}, m_{j}^{(R)},j^{(M)}, m_{j}^{(M)}, \mathbf{r})$ specifies a deterministic trajectory in terms of the states $(j^{(R)}, m_{j}^{(R)})$ and $(j^{(M)}, m_{j}^{(M)})$ of the internal (spin) degrees of freedom of the reservoir and memory-ancilla system, respectively, and the coordinates $\mathbf{r}$ associated with the spatial degrees of freedom, $H_{\rm ext}^{(T)}$ is the Hamiltonian associated with the external (spatial) degrees of freedom of the total reservoir-memory-ancilla system, and $Z_{I}$ and $Z_{E}$ are the respective partition functions.
The Lagrange multiplier $\beta$ is the inverse temperature of the external degrees of freedom of the combined reservoir-memory-ancilla system and $\gamma$ and $\lambda$ are the initial inverse ``spin temperatures'' of the reservoir and memory, respectively.  As mentioned in the main text, we assume the reservoir is sufficiently large (i.e. $N\gg 1$) that erasing one bit of information changes $\gamma$ by a negligible amount.
Each trajectory is labelled uniquely by its initial point $\mathbf{z}_0$, i.e. $\mathbf{z}=\mathbf{z}(\mathbf{z}_0,t)$ where $\mathbf{z}_0=\mathbf{z}(\mathbf{z}_0,t_0)$ at $t=t_0$.
Although the spatial degrees of freedom enable the collisions to take place, the kinetic energy associated with them does not contribute to the cost of the erasure processes because in the absence of an external magnetic field (which is the situation we consider), the spatial and spin degrees of freedom are decoupled.

The traditional Jarzynski equality relates the average exponentiated work with the change in free energy.
In our case we are interested in the average exponentiated spinlabor, i.e.
\bea
\langle e^{-\gamma\mathcal{L}_{s}}\rangle=\sum_{\mathbf{z}} f(\mathbf{z},t) e^{-\gamma\hbar(\Delta m^{(R)}_{j}+\Delta m^{(M)}_{j})}, \label{eqn:expect}
\eea
with respect to the Lagrange multiplier $\gamma$ for the reservoir.
The spinlabor is assumed to be done over a time interval from $t'_0$ to $t$, and the symbol $\Delta m_j^{(\cdot)}$ represents the change
\begin{align}   \label{eqn:Delta m_j}
    \Delta m_j^{(\cdot)}\equiv m_j^{(\cdot)}(t)-m_j^{(\cdot)}(t'_0)
\end{align}
over the deterministic trajectory $\mathbf{z}(\mathbf{z}_0,t)$, where $m_j^{(\cdot)}(t)$ represents the corresponding value of the $z$ component of spin angular momentum at time $t$.

There is a natural division in the erasure protocol between
\begin{list}{(\arabic{enumii})}{\usecounter{enumii}}
\item the first CNOT operation on the memory-ancilla system, and
\item the remainder of the erasure process,
\end{list}
because (1) is associated with the Lagrange multiplier $\lambda$ of the memory-ancilla system whereas (2) involves the equilibration of the memory-ancilla system with the reservoir and is, therefore, associated with the Lagrange multiplier $\gamma$.
Hence, we rewrite Eq. \eqref{eqn:expect} as follows
\begin{align}
\langle e^{-\gamma\mathcal{L}_{s}}\rangle&=\langle e^{-\gamma\mathcal{L}_{s}^{(1)}-\gamma\mathcal{L}_{s}^{(2)}}\rangle\nonumber \\ &=\langle e^{-\gamma\mathcal{L}_{s}^{(1)}}\rangle \langle e^{-\gamma\mathcal{L}_{s}^{(2)}}\rangle,  \label{eqn:product of exp spinlabor}
\end{align}
where $\mathcal{L}_{s}^{(1)}$ is the spinlabor incurred by the first CNOT operation and $\mathcal{L}_{s}^{(2)}$ is the spinlabor incurred by the CNOT operations in the remainder of the erasure process.
The values of $t'_0$ and $t$ in \eq{eqn:Delta m_j} will be taken to correspond to the starting and ending times of each part.
The expectation value factorizes into separate expectation values in the second line of \eq{eqn:product of exp spinlabor} because the equilibration of the memory-ancilla system with the reservoir ensures that costs $\mathcal{L}_{s}^{(1)}$ and $\mathcal{L}_{s}^{(2)}$ are uncorrelated.

For part (1), there is no change to the reservoir (as equilibration has not yet occurred) and so  $\Delta m^{(R)}_{j}=0$.
Thus, the corresponding expectation value can be written as
\bea   \label{eqn:part 1 as sum over prob}
\langle e^{-\gamma\mathcal{L}_{s}^{(1)}}\rangle=\sum_{m^{(M)}_{j}} P(m^{(M)}_{j}) e^{-\gamma\hbar\Delta m^{(M)}_{j}},
\eea
where $P(m^{(M)}_{j})\equiv\sum_{\mathbf{z}=m^{(M)}_j}f(\mathbf{z})=e^{-\lambda\hbar m^{(M)}_{j}}\!\!/Z_{M}$.
In deriving this result we have made use of the fact that any degree of freedom that does not appear in the expression $e^{-\gamma\mathcal{L}_{s}^{(1)}}$ will be traced over.  This is the reason the spatial degree of freedom, for example, does not appear explicitly in \eq{eqn:part 1 as sum over prob}.
Consider the case where the memory spin is initially completely mixed which means that
%We choose initial condition for the stage one as below
%
%\bea
%P(y,0)=
%\begin{cases}
%\frac{1}{2}, & m=0\\
%\frac{1}{2}, & m=1
%\end{cases}
%\eea
%
%\bea
%P(y,t)=
%\begin{cases}
%\frac{1}{2}, & m=0\\
%\frac{1}{2}, & m=2.
%\end{cases}
%\eea
there are two equally-likely outcomes for the CNOT operation: either the ancilla spin remains in the $|{\downarrow}\rangle$ state and so $\Delta{m_{j}}=0$, or the ancilla spin is flipped to $|{\uparrow}\rangle$ and so $\Delta{m_{j}}=1$.
Evaluating \eq{eqn:part 1 as sum over prob} for this case then gives
\begin{align}  \label{eqn:part 1}
\langle e^{-\gamma\mathcal{L}_{s}^{(1)}}\rangle&=\frac{1}{2} e^{-\gamma 0 \hbar}+\frac{1}{2} e^{-\gamma 1 \hbar}\nonumber \\&=\frac{1+e^{-\gamma\hbar}}{2}.
\end{align}

For part (2) of the erasure process we have
\begin{align*}
    \langle e^{-\gamma\mathcal{L}_{s}^{(2)}}\rangle&=\sum_{\mathbf{z}} f(\mathbf{z}, t) e^{-\gamma\hbar(\Delta m^{(R)}_{j}+\Delta m^{(M)}_{j}) }
\end{align*}
where
\bea
   f(\mathbf{z},t)=\frac{e^{-\beta H_{\rm ext}^{(T)}-\gamma\hbar [m_{j}^{(R)}(t)+ m_{j}^{(M)}(t)]}}{Z_{I} Z_{E}}.
\eea
Louiville's theorem \cite{Jarzynski1999} implies
\begin{align*}
        f(\mathbf{z},t)&=f(\mathbf{z}_0,t'_0)\\
        &=\frac{e^{-\beta H_{\rm ext}^{(T)}-\gamma\hbar [m_{j}^{(R)}(t'_0)+ m_{j}^{(M)}(t'_0)]}}{Z_{I}^{(i)} Z_{E}^{(i)}}
\end{align*}
and so
\begin{align}
    \langle e^{-\gamma\mathcal{L}_{s}^{(2)}}\rangle&=\sum_{\mathbf{z}}
    \frac{e^{-\beta H_{\rm ext}^{(T)}-\gamma\hbar [m_{j}^{(R)}(t)+ m_{j}^{(M)}(t)]}}{Z_{I}^{(i)} Z_{E}^{(i)}}\nonumber\\
&=\frac{Z_{I}^{(f)}}{Z_{I}^{(i)}}, \label{eqn:partition}
\end{align}
where the superscript $(i)$ and $(f)$ label initial and final values, respectively, and we have made use of the fact that $Z_{E}^{(f)}=Z_{E}^{(i)}$.
Although the spatial degrees of freedom do not contribute to the erasure cost, including them in $\mathbf{z}$ makes the trajectory deterministic and this enables  the application of Liouville theorem.

Our assumption that the erasure of 1 bit only changes the Lagrange multiplier $\gamma$ associated with the reservoir by a negligible amount implies that the partition function of the reservoir correspondingly also changes by a negligible amount. Thus $Z_{I}^{(f)}/Z_{I}^{(i)}$ is equal to the ratio of the partition functions of just the memory-ancilla system.  At the beginning of part (2) of the protocol (i.e. at $t=0$) the memory-ancilla system is described by the probability distribution given in \eqs{eqn:1} and \eqr{eqn:2} with $m=1$, whereas at the end of the erasure process all memory-ancilla spins are in the state $\ket{\dn}$ (representing complete erasure). Calculating the corresponding values of the partition function then yields
\bea  \label{eqn:part 2}
\langle e^{-\gamma\mathcal{L}_{s}^{(2)}}\rangle=\frac{1}{1+e^{-2\gamma\hbar}}.
\eea
Substituting the results \eqs{eqn:part 1} and \eqr{eqn:part 2} into \eq{eqn:product of exp spinlabor} gives
\bea   \label{eqn:jarzynski-like equality}
\langle e^{-\gamma\mathcal{L}_{s}}\rangle=\frac{1+e^{-\gamma\hbar}}{2(1+e^{-2\gamma\hbar})}\ .
\eea
We refer to this as our \emph{Jarzynski-like} equality because it is an equality for spinlabor in the VB erasure scheme in analogy to Jarzynski's equality \cite{Jarzynski1999} for work and free energy.

%We can also prove a general fluctuation theorem for generalized Gibbs ensembles for applications other than erasure, but we do not present it here as it is beyond the scope of this manuscript and will be discussed elsewhere.

\section{Jarzynski-like Bound}
We now use \eq{eqn:jarzynski-like equality} to analyze the fluctuations in the spinlabor cost in the same way that Jarzynski analyzed the fluctuations in the work-free energy relation \cite{Jarzynski1999}.
Consider the following expression, which is twice the right side of \eq{eqn:expect},
\bea
   A &=& \sum_{\mathbf{z}}f(\mathbf{z},t) e^{-\gamma\hbar(\Delta m^{(R)}_{j}+\Delta m^{(M)}_{j})+\ln 2} \nonumber \\
   &=& \sum_{x}Pr(x\hbar) e^{-\gamma\hbar x+\ln 2}
   \label{eqn:definition of A}
\eea
where, for convenience, we let $x=\Delta m^{(R)}_{j}+\Delta m^{(M)}_{j}$ and define $Pr(x\hbar)\equiv\sum_{\mathbf{z}=x}f(\mathbf{z})$.
By conservation of spin angular momentum, $x\hbar$ is equal to the spinlabor $\mathcal{L}_s$ done on the memory-ancilla system by the CNOT operation, and so $Pr(\mathcal{L}_s)$ represents the probability that the cost of the erasure is $\mathcal{L}_s$.
As the summand in \eq{eqn:definition of A} is positive, restricting the sum to values of $x$ that satisfy $-\gamma x\hbar+\ln 2 \geq \gamma\epsilon$ for $\epsilon>0$ gives
\bea \label{eqn:inequality}
   \sum_{-\gamma x\hbar+\ln 2\geq \gamma\epsilon }\kern-10pt Pr(x\hbar)e^{\gamma \epsilon } \leq \sum_{x}Pr(x\hbar)e^{\gamma \epsilon} \leq A.
\eea
We represent the probability that the cost $x\hbar$ violates the VB bound [Eq.~(3) in the main text] by $\epsilon$ as
\bea
   Pr(-\gamma x\hbar+\ln 2 \geq \gamma\epsilon )\equiv \kern-5pt\sum_{-\gamma x\hbar+\ln 2 \geq \gamma\epsilon }\kern-10pt Pr(x\hbar) \label{eqn:probchange}
\eea
and so from \eq{eqn:inequality} we have
\bea
    Pr(-\gamma x\hbar+\ln 2 \geq \gamma\epsilon)e^{\gamma \epsilon } \leq A .
\eea
Thus the probability that the spinlabor cost $x\hbar=\mathcal{L}_s$ violates VB's bound by $\epsilon$ satisfies
\bea   \label{eqn:bound A expanded}
Pr(\mathcal{L}_s \leq \gamma^{-1} \ln 2-\epsilon) &\leq& A e^{-\gamma \epsilon }
\eea
where $A$, being twice the right side of \eq{eqn:expect}, is found from \eq{eqn:jarzynski-like equality} to be
\begin{align}
        A=\frac{1+e^{-\gamma\hbar}}{1+e^{-2\gamma\hbar}}\ .
\end{align}
It is convenient to define the probability of violation more compactly as
\begin{align}   \label{eqn:compact prob of violation}
        Pr^{(v)}(\epsilon)\equiv {Pr}(\mathcal{L}_s\leq \gamma^{-1} \ln{2} -\epsilon )
\end{align}
and hence, from \eq{eqn:bound A expanded},
\bea   \label{eqn:bound A}
    Pr^{(v)}(\epsilon) &\leq& A e^{-\gamma \epsilon } .
\eea

\newpage
\section{Tighter Bound}
The probability of violation can be bounded tighter than \eq{eqn:bound A} by restricting the sum in \eq{eqn:definition of A}.  In particular, setting
\bea   \label{eqn:definition of B}
  B\equiv \sum_{-\gamma x\hbar +\ln 2\geq 0} Pr(x\hbar) e^{-\gamma x\hbar+\ln 2}
\eea
and following a similar argument to the one in the previous section with \eq{eqn:definition of B} in place of \eq{eqn:definition of A} gives
\bea   \label{eqn:bound B}
  Pr^{(v)}(\epsilon) \leq B e^{-\gamma\epsilon}.
\eea
As $B\leq A$, \eq{eqn:bound B} bounds the probability of violation tighter than \eq{eqn:bound A}.

\section{Semi-analytic bound}

An alternate way to estimate the probability of violation $Pr^{(v)}(\epsilon)$ is to fit an exponentially decaying function to it at  $\epsilon=0$ and $\epsilon=\hbar$ using the analytical solution in \eq{eqn:analy} where, according to \eq{eqn:compact prob of violation},
\begin{align}   \label{eqn:probvio}
        Pr^{(v)}(\epsilon)=\sum_{q=0}^{b-\epsilon/\hbar}P_\infty(q) ,
\end{align}
and we select only particular values of $\gamma$ given by
\begin{align}   \label{eqn:special values of gamma}
       \gamma=\frac{\ln(2)}{b\hbar}
\end{align}
for $b=1,2,3,\ldots$.  Selecting these specific values of $\gamma$ ensures that the VB bound given in Eq.~(3) in the main text is a multiple of $\hbar$, i.e. $b\hbar=\ln(2)/\gamma$.
More specifically, we approximate $Pr^{(v)}(\epsilon)$ by the function
\begin{align}  \label{eqn:fitted function}
             \widetilde{Pr}{}^{(v)}(\epsilon)=C e^{-a\epsilon}
\end{align}
where the amplitude and decay parameters, $C$ and $a$, are to be determined as follows.  The value of $C$ is determined by requiring $\widetilde{Pr}{}^{(v)}(\epsilon)=Pr^{(v)}(\epsilon)$ for $\epsilon=0$, and so
\begin{align}   \label{eqn:bound C}
     C=Pr^{(v)}(0) .
\end{align}
The value of the decay rate $a>0$ is determined by setting $\widetilde{Pr}{}^{(v)}(\epsilon)=Pr^{(v)}(\epsilon)$ for $\epsilon=\hbar$, i.e.
\begin{align}  \label{eqn:second}
             Pr^{(v)}(0) e^{-a\hbar}=Pr^{(v)}(\hbar) .
\end{align}
Using Eq.~\eqref{eqn:probvio} to replace the right side gives
\bea
   Pr^{(v)}(0) e^{-a\hbar}
   &=& \sum_{q=0}^{b-1} P_\infty(q) = \sum_{q=0}^{b} P_\infty(q)-P_\infty(b)\nonumber \\
   &=& Pr^{(v)}(0)-P_\infty(b) \label{eqn:rearr}
\eea
which, on rearranging, becomes
\bea
     \frac{1}{1-e^{-a\hbar}}=\frac{Pr^{(v)}(0)}{P_\infty(b)} ,
\eea
and, on solving for $a$, yields the analytical result for the decay rate as
\begin{align} \label{eqn:decay rate a}
    a=-\frac{1}{\hbar} \ln\left\{1-\left[\frac{Pr^{(v)}(0)}{P_\infty(b)}\right]^{-1}\right\} .
\end{align}
In order to calculate it, we need to evaluate the expression in $[\ldots]$ brackets. Using the fact that, from Eq.~\eqref{eqn:probvio},
\begin{align*}
        Pr^{(v)}(0)=P_\infty(b)+P_\infty(b-1)+\ldots+P_\infty(0)
\end{align*}
we find
\begin{align}
    \frac{Pr^{(v)}(0)}{P_\infty(b)}=\frac{P_\infty(b)}{P_\infty(b)}+\frac{P_\infty(b-1)}{P_\infty(b)}+.\;.\;.\;+\frac{P_\infty(0)}{P_\infty(b)}. \label{eqn:theterms}
\end{align}
The terms on the right side can be expanded using  Eq.~\eqref{eqn:analy}. For example,
\bea
    \frac{P_\infty(b-1)}{P_\infty(b)}=\frac{1-r^{b}}{r^{b}}\frac{1-r^{b-1}(1-r)}{1-r^{b}(1-r)}
\eea
and, noting that \eq{eqn:define r} and \eq{eqn:special values of gamma} imply $r^{b}=\frac{1}{2}$, we find
\bea
          \frac{P_\infty(b-1)}{P_\infty(b)}= \frac{2-r^{-1}(1-r)}{1+r}.
\eea
The next two terms are
\begin{align*}
      \frac{P_\infty(b-2)}{P_\infty(b)}&=(2r-1) \frac{2-r^{-2}(1-r)}{1+r}\\
     \frac{P_\infty(b-3)}{P_\infty(b)}&=(2r-1)(2r^{2}-1) \frac{2-r^{-3}(1-r)}{1+r} .
\end{align*}
Continuing in this way we find
\begin{align*}
         \frac{P_\infty(b-n)}{P_\infty(b)}&=(2r-1)(2r^{2}-1)\ldots (2r^{n-1}-1) \\
         &\qquad \times\frac{2-r^{-n}(1-r)}{1+r}
\end{align*}
for $n$ being a positive integer less than $b$, and
\begin{align*}
       \frac{P_\infty(0)}{P_\infty(b)}&=(2r-1)(2r^{2}-1)\ldots (2r^{b-1}-1)\frac{2r}{1+r}.
\end{align*}
The right side of \eq{eqn:theterms} can easily be evaluated numerically using these results, and the outcome can then be used to find the value of the decay rate $a$ in \eq{eqn:decay rate a} for the specific values of $\gamma$ given in \eq{eqn:special values of gamma}.
The resulting approximation given by $\widetilde{Pr}{}^{(v)}(\epsilon)$ in \eq{eqn:fitted function} equals $Pr^{(v)}(\epsilon)$ for $\epsilon=0$ and $\epsilon=\hbar$ (by construction) and is found, numerically, to upper bound $Pr^{(v)}(\epsilon)$ for $\epsilon>\hbar$.
Hence, we find semi-analytically that
\begin{align*}
     Pr^{(v)}(\epsilon) \leq C e^{-a \epsilon } .
\end{align*}

\begin{figure}%[ht]
\centering
%\captionsetup{justification=RaggedRight}
\vspace{-2mm}
      \includegraphics[width=0.40\textwidth]{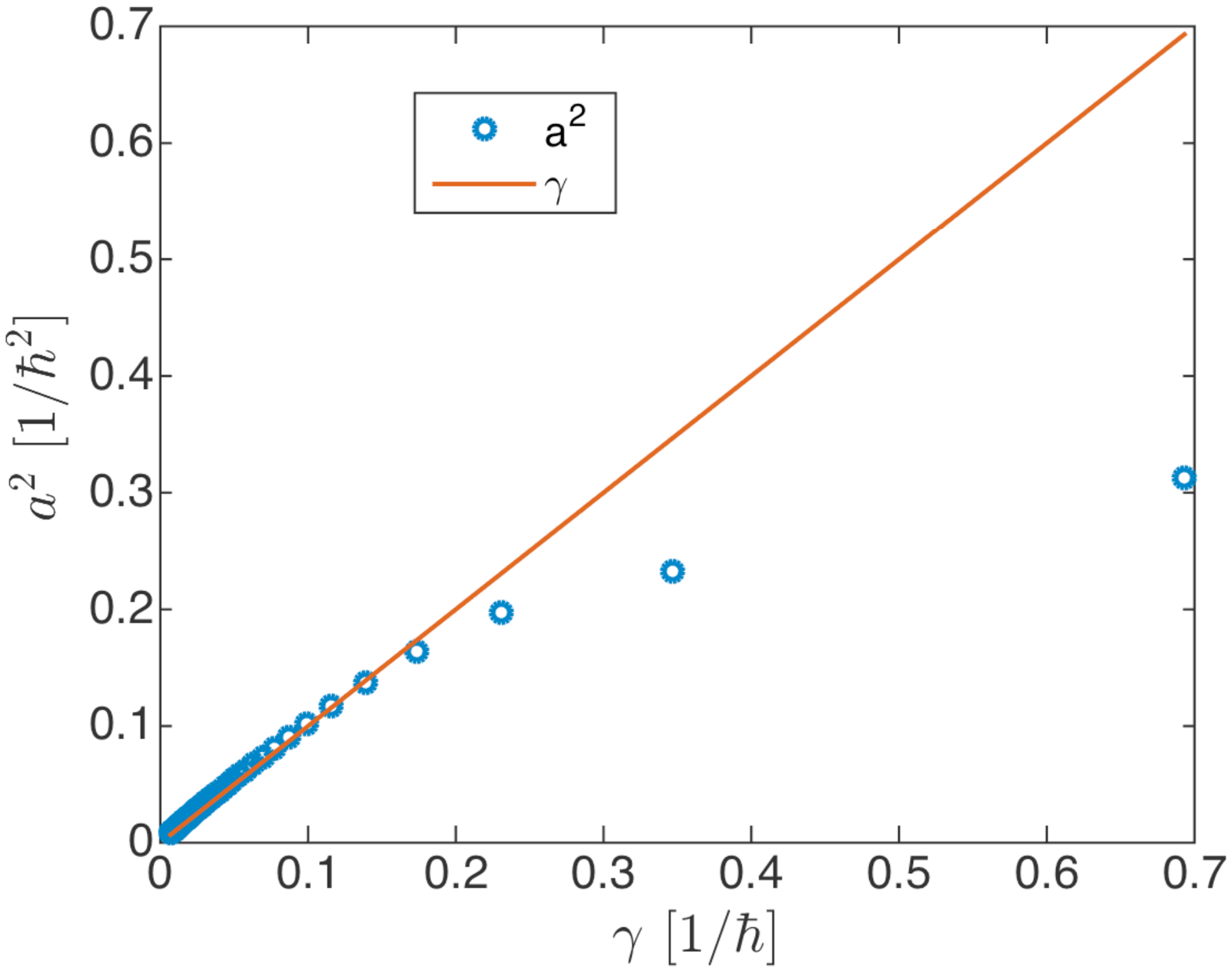}
  \vspace{-12mm}
\caption{The behaviour of the exponential decay rate $a^2$ in the limit $\gamma\to 0$.  The blue circles represent values of $a^2$ plotted as a function of $\gamma$ for the particular values of $\gamma$ given by \eq{eqn:special values of gamma}.  As $\gamma\to 0$ the blue circles approach the orange line which represents values of $\gamma$, and thus illustrates graphically that $\lim_{\gamma\to 0} a=\sqrt{\gamma/\hbar}$.  }
  \label{fig:fittedbound}
  \vspace{-2mm}
\end{figure}

Moreover, we find a simple expression for $a$ in the limit $\gamma\to 0$ as follows.  In Fig.~ \ref{fig:fittedbound} we plot $a^2$ as a function of $\gamma$ to show that $a^2$ approaches $\gamma$ as $\gamma\to 0$ which suggests the quite remarkable result that
\begin{align*}
        \lim_{\gamma\to 0} a = \sqrt{\frac{\gamma}{\hbar}} .
\end{align*}
We have found numerically that the function, $\widetilde{Pr}{}^{(v)}(\epsilon)$ in \eq{eqn:fitted function}, that corresponds to this limiting value of $a$ upper bounds the probability of violation,  $Pr^{(v)}(\epsilon)$,  for $\epsilon> 0$, i.e.
\bea  \label{eqn:semi-analytic bound}
     {Pr}^{(v)}(\epsilon )\leq C e^{-\sqrt{\frac{\gamma}{\hbar}}\epsilon},
\eea
where $C$ is given by \eq{eqn:bound C}.  The fact that \eq{eqn:semi-analytic bound} gives a tighter bound than \eqs{eqn:bound A} and \eqr{eqn:bound B} is illustrated in Figs.~1(b) and 2(b) of the main text.  We refer to \eq{eqn:semi-analytic bound} as a \emph{semi-analytical} bound on the probability of violation given the combination of analytical and numerical methods we used in deriving it.

\bibliography{Supplimaterial}